\documentclass[aps,prd,showpacs,showkeys,amsmath,amssymb,
twocolumn,
floatfix,
superscriptaddress，
xcolor
]{revtex4-1}

\usepackage{graphicx}
\usepackage{times}
\usepackage{multirow}
\usepackage{verbatim}
\usepackage{color,soul}
\usepackage{xcolor}
\usepackage{cancel}
\usepackage{graphicx}
\usepackage{subcaption}
\usepackage[normalem]{ulem}
\graphicspath{{figures/}}

\def\U {{}^{235}\text{U}}
\def\Um{{}^{238}\text{U}}
\def\Th{{}^{232}\text{Th}}
\def\K{{}^{40}\text{K}}

\begin{document}


\title{New calculation of the geo-neutrino energy spectrum and its implication}
\author{Yu-Feng Li}
\email{liyufeng@ihep.ac.cn}
\affiliation{Institute of High Energy Physics,
Chinese Academy of Sciences, Beijing 100049, China}
\affiliation{School of Physical Sciences, University of Chinese Academy of Sciences, Beijing 100049, China}

\author{Zhao Xin}
\email{xinzhao@ihep.ac.cn}
\affiliation{Institute of High Energy Physics,
Chinese Academy of Sciences, Beijing 100049, China}
\affiliation{School of Physical Sciences, University of Chinese Academy of Sciences, Beijing 100049, China}

\date{\today}

\begin{abstract} 
The energy spectrum of geo-neutrinos plays a vital role in the experimental measurement of geo-neutrinos that have profound implications for both particle physics and earth sciences. In this letter, we present a state-of-the-art calculation of the energy spectrum of geo-neutrinos originating from the beta decay of Uranium-238 and Thorium-232. Our calculation is underpinned by the latest updates in the nuclear database, accounts for previously overlooked forbidden transitions, and incorporates advanced corrections for the beta decay. This brand new geo-neutrino flux model, compared to the widely-used estimates from Enomoto~\cite{Enomoto}, reveals notable distinction in the energy spectrum shape because of our comprehensive approach. When considering the inverse beta decay (IBD) detection process, our findings show a significant deviation in the predicted IBD yield of around 4\% for Uranium-238 and 9\% for Thorium-232 decay chains. The implications of using the new geo-neutrino flux model for the experimental analysis are substantial, potentially affecting the analysis results of geo-neutrino measurements of KamLAND and Borexino by around 10\% to 20\%.
Our study represents a significant advancement in geo-neutrino research, establishing a new benchmark for accuracy and reliability in the field.

\end{abstract}

\maketitle

Geo-neutrinos generated from the beta decay chains of the nuclear isotopes with long half-lives such as $\Um$, $\Th$, and $\K$ carry the Earth's internal information which helps us to understand key problems of the geophysical and geochemical sciences~\cite{Eder1966,Javoy1997, Nunokawa:2003dd, Marx:1969be,Ruedas:2017dgt}. During these decay processes, both antineutrinos and energies are released, with the latter known as radiogenic heat, which contributes to the Earth's overall heat production. Thus these elements are commonly known as Heat-Producing Elements (HPEs)~\cite{Sramek:2012nk}. The production of antineutrinos and radiogenic heat is intrinsically linked, enabling the use of geo-neutrino measurements as a tool for estimating the Earth's inaccessible radiogenic heat production~\cite{Fiorentini:2002bp, Fiorentini:2005mr, Dye:2011mc, McDonough:2019ldt}.

The detection of geo-neutrinos has implications for geosciences and particle physics collectively and individually~\cite{Bellini:2021sow}. Particle physicists are interested in understanding the nature of these particles, while the geological community seeks transformational insights from the detection of geo-neutrinos, particularly in accurately determining the global inventory of HPEs~\cite{Sramek:2012nk}, and these elements account for over $99\%$ of the radiogenic heat production in the Earth and, together with the primordial energy of accretion and core segregation, define the total power budget of the planet~\cite{Domogatsky:2004be}. Current and future liquid-scintillator detectors will be able to directly measure the amount of geo-neutrinos from $\Um$ and $\Th$ in the Earth~\cite{Rothschild:1997dd}, but not the signal from $\K$, as the antineutrino energy from the beta decay of $\K$ is below the threshold energy of the inverse beta decay (IBD) reaction ($\bar{\nu}_e + p \to e^+ + n$)~\cite{Ricciardi:2022pru, Strumia:2003zx}.

The summation method, also known as the \textit{ab-initio} method~\cite{King:1958zz,PhysRev.170.931,Vogel:1980bk,TENGBLAD1989136}, is often applied to the determination of the geo-neutrino energy spectrum. 
In this approach, the entire energy spectrum is calculated by directly adding up each individual beta branch, with each one being weighted according to its beta decay activity.
The summation method, developed for the precision prediction of reactor antineutrino spectrum, was first introduced in Ref.~\cite{King:1958zz} and subsequently underwent sequential refinements and revisions, as documented in Refs.~\cite{PhysRev.170.931,Vogel:1980bk, TENGBLAD1989136}.
This approach depends entirely on the available nuclear databases~\cite{jeff_web,endf_web, ensdf_web}. The currently widely used geo-neutrino energy spectrum is evaluated in 2005 in Ref.~\cite{Enomoto}. 
One may immediately question whether updates to the nuclear database over approximately 20 years would significantly impact the expected energy spectrum. Additionally, in Ref.~\cite{Enomoto}, all branches are treated as allowed transitions, and high-order corrections for each decay branch are not considered. Therefore, an updated calculation of the geo-neutrino energy spectrum, employing the latest database and an advanced description of the beta spectrum, is essential and will play a crucial role in the precision measurements of geo-neutrinos.

In this Letter, we present a new assessment of individual geo-neutrino fluxes by using the summation method from the decay chains of two dominant isotopes with long half-lives, namely, i.e., $\Um$ and $\Th$. The integration of the updated nuclear database, consideration of forbidden transitions, and inclusion of higher-order corrections yield significant spectral deviation compared to the currently commonly used fluxes in Ref.~\cite{Enomoto}. This update significantly influences the expected geo-neutrino event rates in liquid-scintillator detectors, resulting in relative differences of approximately 4\% and 9\% for the $\Um$ and $\Th$ decay chains, respectively.
Very recently, the KamLAND~\cite{KamLAND:2022vbm} and Borexino~\cite{Borexino:2019gps} collaborations have published their latest results on the geo-neutrino observation, with data collected over approximately 5227 and 3263 days, respectively.
In the analysis of geo-neutrino measurements conducted by both collaborations, the fluxes reported by Enomoto have been utilized to present their fitting results. To assess the impact of the newly developed geo-neutrino flux model, we incorporate these updated fluxes into the analysis of data published by both collaborations, which result in substantial effects on the fit results. The remainder of this letter will outline the principal findings of the new geo-neutrino flux model and its implications for data analysis. More technical details are provided in the Appendix.


The geo-neutrino flux and energy spectrum are calculated using the summation method, which involves evaluating the energy spectrum of a single beta decay along with the relevant information from nuclear database. For the single beta decay calculation, a universal description is employed, as outlined in Ref.~\cite{Fermi:1934hr}, and can be expressed as follows:
\begin{align}
    S_\nu (E_\nu)= & K p_\nu E_\nu (E_0 - E_\nu)^2 F(Z, E_\nu)  \nonumber \\ 
    & C(Z, E_\nu) \left[1+ \delta(Z,A,E_\nu) \right],
    \label{eq.single_branch}
\end{align}
where $K$ is the normalization factor, and $F(Z, E_\nu)$ is the Fermi function that accounts for the Coulomb interactions between between the emitted electron and daughter nucleus. 
The shape factor $C(Z, E_\nu)$ characterizes the distortion in the forbidden decay relative to the allowed Gamow-Teller (GT) transition. The form of this function is based on the description provided in Ref.~\cite{Li:2019quv} and is presented in Table~\ref{tab.shape_factors}. The additional higher-order corrections $\delta (Z,A,E_\nu)$ can be broken down into three components: the radiative correction (RC) $\delta_{\rm{RC}}$~\cite{Sirlin:1967zza,Sirlin:2011wg}, the finite size (FS) correction $\delta_{\rm{FS}}$~\cite{Hayes:2013wra,Wang:2016rqh}, and the weak magnetism (WM) correction $\delta_{\rm WM}$~\cite{Hayes:2013wra}. More detailed information regarding the calculation of the beta decay spectrum is provided in the Appendix.

Using data from the Evaluated Nuclear Structure Data Files (ENSDF)~\cite{ensdf_web} related to the decay chain of a specific nuclide, we can directly calculate the total neutrino spectrum for this decay chain as follows:
\begin{align}
    S_{X} = \sum_{ij} R_{ij} \sum_k I_{ij,k} S_\nu^{ij,k},
\end{align}
where $S_{X}$ represents the geo-neutrino spectrum arising from the decay chain of the nuclide $X$ ($X=\Um\,\mathrm{or}\,\Th$), $R_{ij}$ denotes the weight of the branching ratio in the decay process from isotope $i$ to isotope $j$, and $I_{ij,k}$ refers to the $k$-th branch ratio for the transition from isotope $i$ to isotope $j$. These values are detailed in Table~\ref{tab.beta_decay_in_238U} and Table~\ref{tab.beta_decay_in_232Th}.


In the process of $\Um$ $\to$ ${}^{206}\rm{Po}$, there are eight $\alpha$ decays and six $\beta$ decays. Among these decays, only three nuclides, ${}^{234}\rm{Pa}$, ${}^{214}\rm{Bi}$ and ${}^{210}\rm{Tl}$, can produce antineutrinos above the IBD energy threshold of 1.806 MeV, with the contribution from ${}^{210}\rm{Tl}$ being negligible.
In the decay chain of $\Th$ $\to$ ${}^{208}\rm{Pb}$, there are six $\alpha$ decays and four $\beta$ decays. Only the decays from ${}^{228}\rm{Ac}$ and ${}^{212}\rm{Bi}$ contribute to the detectable geo-neutrino signals above the threshold of the IBD process.
Compared to the calculation in Ref.~\cite{Enomoto}, this work identifies 77 new transitions in $\Um$ and 14 new transitions in $\Th$, as summarized in Table~\ref{tab.beta_decay_in_238U} and Table~\ref{tab.beta_decay_in_232Th}, respectively. All these additional transitions occur in the low-energy region and do not contribute to the geo-neutrino signals detected by the IBD reaction. 
In addition to the increase in the number of decay branches, another notable change between these two versions of the ENSDF data is the alteration of $Q$-values for certain key transitions.

\begin{figure}
    \centering
    \begin{subfigure}[b]{0.5\textwidth}
        \centering
        \includegraphics[width=\textwidth]{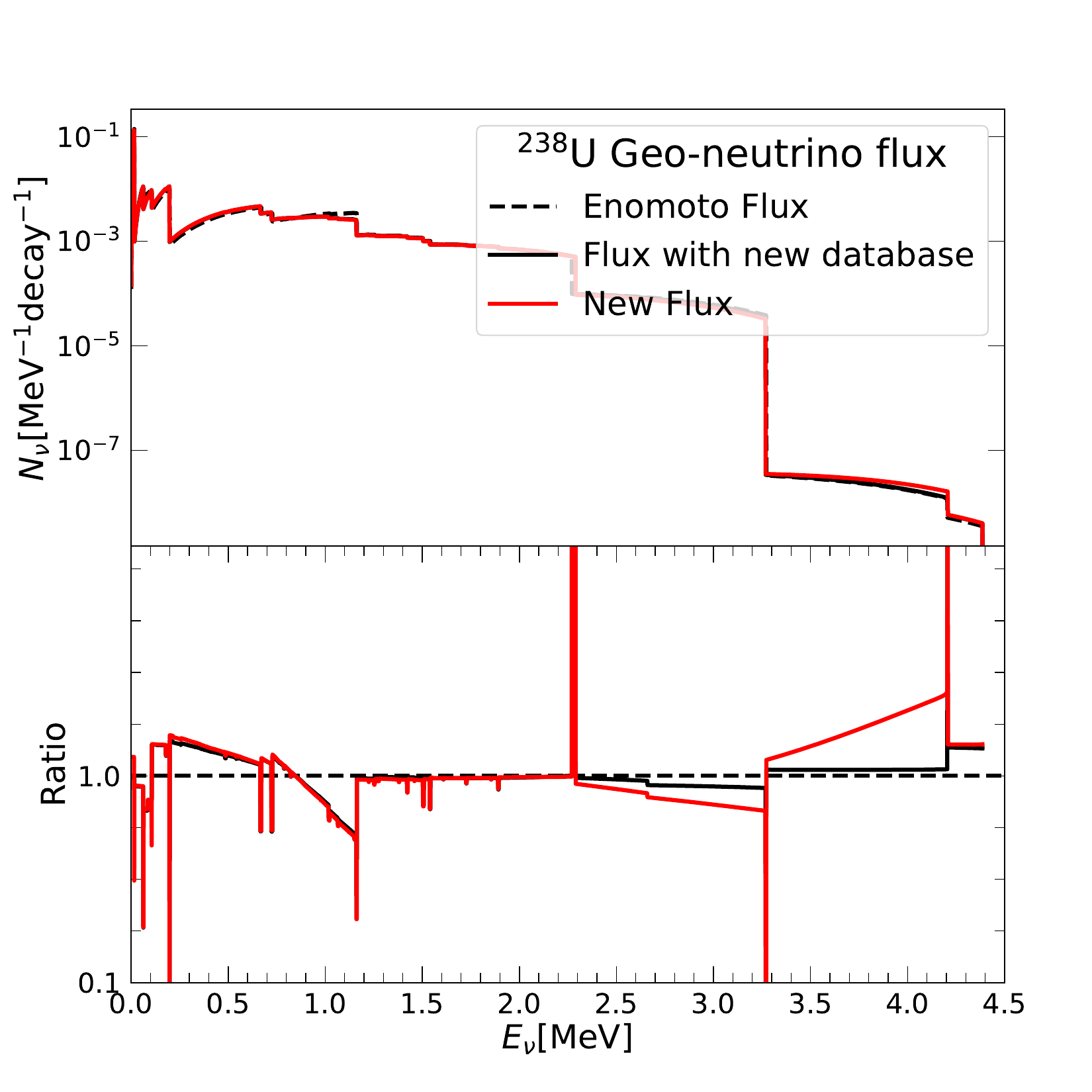}
    \end{subfigure}
    \hfill
    \begin{subfigure}[b]{0.5\textwidth}
        \centering
        \includegraphics[width=\textwidth]{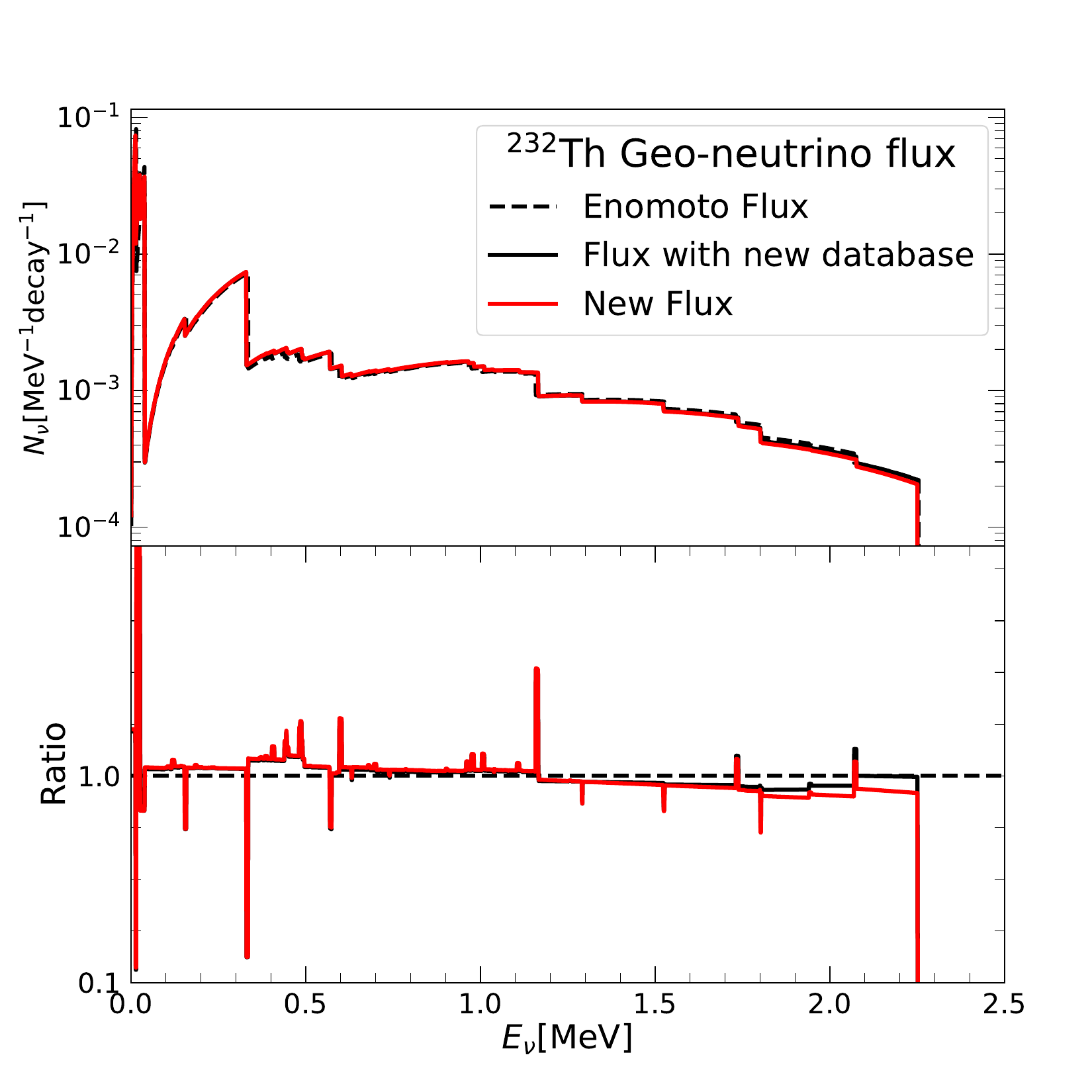}
    \end{subfigure}
    \caption{
The geo-neutrino energy spectra from the decay chains of $\Um$ and $\Th$.
In the upper panels of each subplot, three fluxes are illustrated: the black dashed lines (Enomoto Flux) represent data from Ref.~\cite{Enomoto}, the black solid lines (Flux with new database) correspond to the flux calculated using the new ENSDF database~\cite{ensdf_web}, maintaining the same beta decay description as in Ref.~\cite{Enomoto}. The red solid lines (New Flux) reflect the estimates derived from the new ENSDF database, incorporating higher-order corrections and accounting for forbidden transitions. 
In the lower panels of each subplot, the ratios of all fluxes to the Enomoto flux are presented, highlighting the relative differences resulting from the updated nuclear database and the consideration of forbidden transitions.}
    \label{fig.spectra}
\end{figure}
In Figure~\ref{fig.spectra}, the geo-neutrino energy spectra for $\Um$ and $\Th$ are shown in the upper panels, while the ratios of our new calculations to the reference Enomoto flux are displayed in the lower panels.
The black dashed lines represent the commonly used calculations from Ref.~\cite{Enomoto}, labeled as "Enomoto Flux" The black solid lines are calculated using the updated nuclear database, following the same methodology as Ref.~\cite{Enomoto}, while excluding forbidden transitions and higher-order corrections; these lines are labeled "Flux with new database".
The red solid lines indicate the new flux that incorporate updates to both the nuclear database and the beta decay spectrum calculation, and are denoted as "New Flux".

In the figure, several noticeable spikes are observed, primarily due to variations in $Q$-values for the same transition across different versions of the nuclear database.
Considering the higher-order corrections and shape factors for forbidden transitions, the red lines exhibit noticeable spectral distortions near the endpoints of the decay branches.
In the lower panel of Figure~\ref{fig.spectra}, the red lines fall below the black and blue lines in the high-energy region due to the influence of forbidden transitions.
According to Table~\ref{tab.forbidden_decay}, the transitions with high $Q$-values correspond to $\Delta J^{\pi} = 1^-$, indicating non-unique first forbidden transitions. The exact relativistic calculation of the Dirac wave function employed in this work for estimating these forbidden transitions may result in an underestimation in the high-energy region compared to allowed transitions. Table~\ref{tab.forbidden_decay} summarizes the transitions that exceed the IBD threshold energy.
For $\Um$, the primary forbidden transition is the decay of ${}^{214}\rm{Bi}$, which contributes approximately $47\%$ to the total $\Um$ geo-neutrino spectrum. Although the decay of ${}^{234} \rm{Pa}$ also plays a significant role, accounting for about $39\%$, the spectral shape of the first non-unique forbidden transition ($\Delta J^\pi = 0^-$) is strikingly similar to that of the allowed transitions. In the case of $\Th$, the decay of ${}^{212}{\mathrm{Bi}}$ is the sole contribution from forbidden transitions that exceeds the IBD threshold.


In current geo-neutrino detection efforts utilizing liquid-scintillator detectors, the IBD process is widely employed. As shown in the lower panels of Figure~\ref{fig.spectra}, the updated spectra for both $\Um$ and $\Th$ are smaller than those from Enomoto in the high-energy region, while exhibiting larger values in the low-energy region. Consequently, when considering the IBD cross section, we can deduce that the geo-neutrino signals calculated using the updated spectra are lower than those based on Enomoto's results. Therefore, to quantitatively compare the relative differences in IBD signals estimated using the spectra from Ref.~\cite{Enomoto} and the updated spectra from this work, it is more effective to use the IBD yield, also known as the cross section per decay, which can be expressed as follows:
\begin{align}
    \sigma_X = \int_{E_\nu^{\rm{thr}}}^{E_\nu^{\rm{max}}} dE_\nu S_X (E_\nu) \sigma_{\rm{IBD}}(E_\nu), 
    \label{eq.ibd_yield}
\end{align}
where $E_\nu$ is the neutrino energy, and $S_X(E_\nu)$ is the geo-neutrino flux from the $\beta$ decay chain of isotope $X$ (with $X=\Um \, \rm{or} \, \Th$), as illustrated in the upper panels of Figure~\ref{fig.spectra}. The term $\sigma_{\rm{IBD}}$ represents the IBD cross section, for which we reference calculations from Refs.~\cite{Ricciardi:2022pru,Strumia:2003zx}.

In Table\ref{tab.ibd_yield}, we provide a comparative summary of the IBD yields derived from the updated spectra presented in this work, relative to those from Ref.~\cite{Enomoto}.
The first row indicates the relative differences arising from two versions of the ENSDF database, which are nearly two decades apart. Moreover, the second row incorporates the effects of forbidden transitions and higher-order corrections compared to the first row. 
Regarding the $\Um$ geo-neutrino flux, there is a significant gap around 2.3 MeV, originating from a transition branch in the decay of ${}^{234}\mathrm{Pa}^{\mathrm{m}} \to {}^{234}\mathrm{U}$. The corresponding $Q$-value for this branch has shifted from 2.269 MeV to the current 2.290 MeV, resulting in a 21 keV interval that contributes a $+3.31\%$ relative difference to the IBD yield.
Taking into account all the new database information, a net change of $+0.16\%$ is obtained.
Finally, with the inclusion of the new beta decay spectrum, we achieve a total reduction of $-3.47\%$ in the IBD yield.
In contrast, for $\Th$, the nuclear database indicates a $-3.88\%$ underestimation of the geo-neutrino flux. Furthermore, the inclusion of forbidden transitions and higher-order corrections results in an additional reduction of approximately $-5.12\%$ of the IBD yield.

\begin{table}
    \centering
    \setlength{\tabcolsep}{4mm}{
    \begin{tabular}{ccc}
    \hline\hline
         Model & $\Um$ &$\Th$ \\
         \hline
         Flux with new database & $+0.16 \%$ & $-3.88 \%$\\
         New Flux & $-3.47 \%$ & $-9.00 \%$\\
    \hline\hline
    \end{tabular}    
    \caption{
Comparative summary of the IBD yields derived from the updated spectra in this work, relative to those from the Enomoto Flux. The relative difference in IBD yields is calculated as  
$(S_i-S_0)/{S_0}$, where $S_i$ represents the IBD yield calculated with the 'Flux with new database' in the first row, and the 'New flux' in the second row. Here, $S_0$ is the IBD yield estimated with the Enomoto Flux.
}
    \label{tab.ibd_yield}}
\end{table}

\begin{figure}
    \centering
    \begin{subfigure}[b]{0.5\textwidth}
        \centering
        \includegraphics[width=\textwidth]{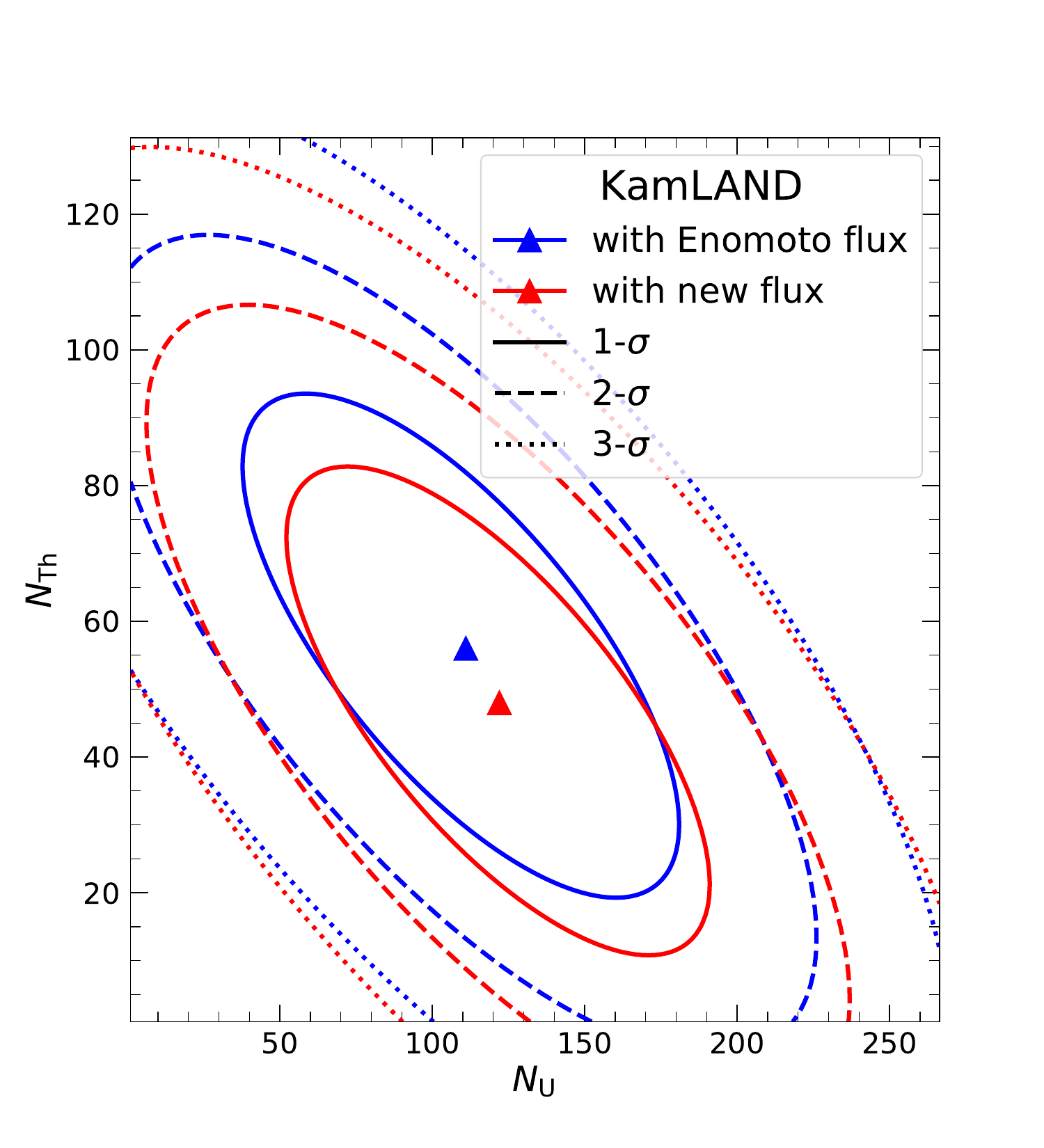}
        \label{fig.KamLAND_fig}
    \end{subfigure}
    \hfill
    \begin{subfigure}[b]{0.5\textwidth}
        \centering
        \includegraphics[width=\textwidth]{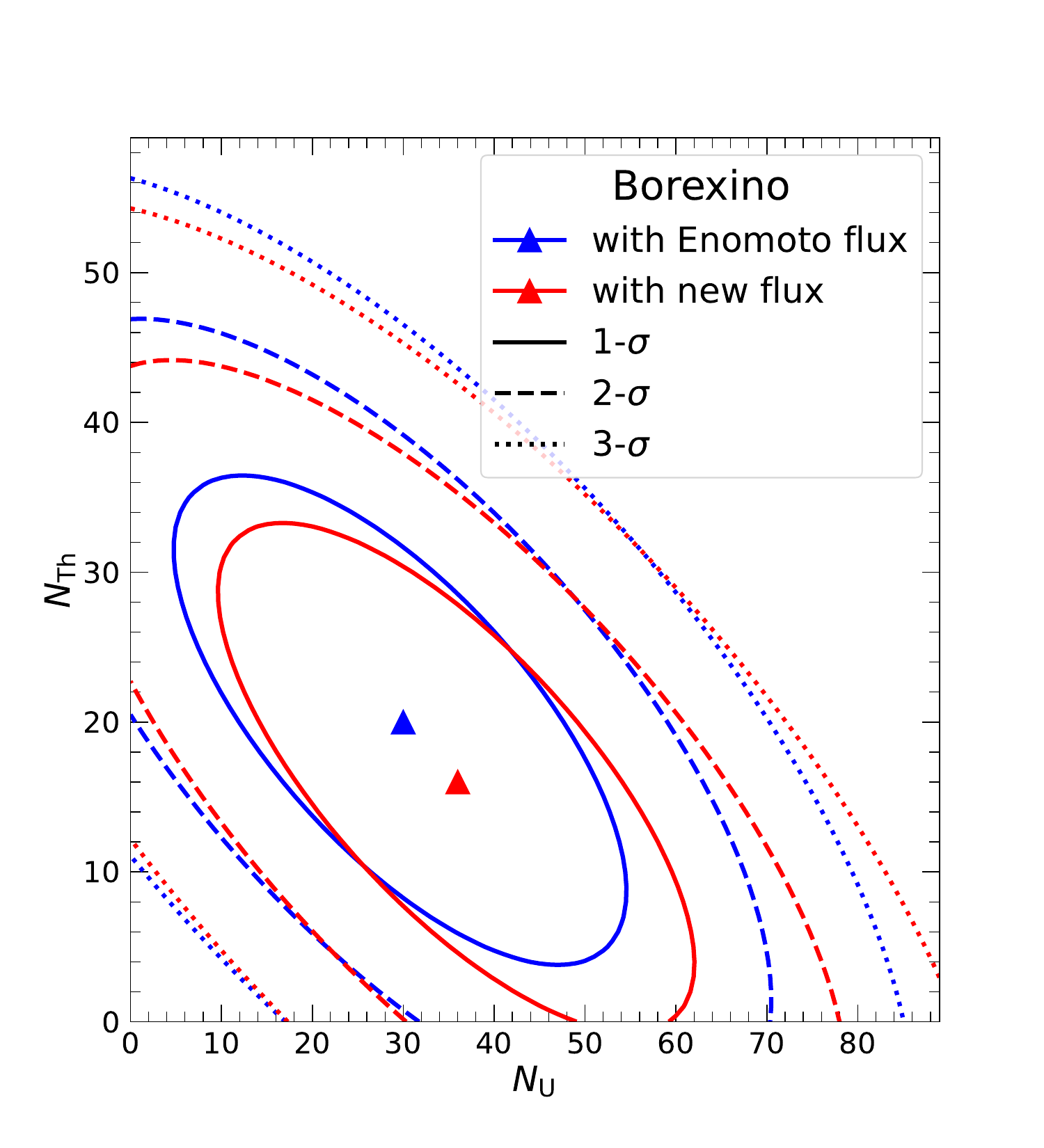}
        \label{fig.Borexino_fig}
    \end{subfigure}
    \caption{
Profiles of $\Delta \chi^2$ for the observed numbers of IBD events from the geo-neutrino contributions of $\Um$ and $\Th$ in KamLAND (upper panel) and Borexino (lower panel). The blue lines and points represent fits based on the Enomoto flux, while the red lines and points are derived from the new flux model.
    }
    \label{fig.analysis}
\end{figure}

The detection of geo-neutrinos is currently a prominent focus of many neutrino experiments~\cite{Araki:2005qa, Chen:2005zza, Borexino:2010dli, Strati:2014kaa, Han:2015roa}. Recently, updated results from geo-neutrino measurements at KamLAND and Borexino were published in Refs.~\cite{KamLAND:2022vbm} and~\cite{Borexino:2019gps}, based on the statistics of 5227 days and 3263 days, respectively. In this work, we first recalculate the results for KamLAND and Borexino using the Enomoto flux~\cite{Enomoto}, which is also employed in the current analyses by both collaborations. Our recalculated results are in agreement with the official releases.
In Table~\ref{tab.fitted_results}, rows 1 through 6 present the results of treating the $\U$ and $\Th$ components as fitting parameters, utilizing data released by the KamLAND and Borexino collaborations in conjunction with the Enomoto flux as inputs.
For KamLAND, our results yield $N_{\mathrm{U}}=111\pm40$ and $N_{\mathrm{Th}}=56\pm30$ when employing the Enomoto flux input.  This corresponds to a total of $167\pm31$ geo-neutrino events, taking into account the correlation coefficient of $r = -0.64$ as derived from Figure~\ref{fig.analysis}. Additionally, we find the Th/U mass ratio to be $7.50\pm4.96$. When the Th/U ratio is fixed at 3.9~\cite{MCDONOUGH1995223}, the total number of geo-neutrino signals increases to $176\pm29$. For Borexino, we obtain $N_{\mathrm{U}}=30\pm16$ and $N_{\mathrm{Th}}=20\pm11$, with a correlation coefficient of $r=-0.72$, resulting in a Th/U mass ratio of $9.83\pm9.88$. This analysis yields $50\pm11$ geo-neutrino events when the Th/U ratio is unconstrained and $52\pm9$ when it is fixed at 3.9. The correlation between the observed $\Um$ and $\Th$ geo-neutrino components can also be seen in Figure~\ref{fig.analysis}. The blue lines in this figure represent the fitting results based on the Enomoto flux input and illustrate the $\Delta \chi^2$ profiles in the $N_{\rm{U}}$-$N_{\rm{Th}}$ plane, analyzing the observed data from both KamLAND and Borexino.


On the other hand, the KamLAND and Borexino data can also be fitted using our updated calculations of geo-neutrino flux inputs. The fitting results are presented from the $7^{\rm{th}}$ to $12^{\rm{th}}$ rows in Table~\ref{tab.fitted_results}.
In comparison to the results based on the Enomoto flux, both KamLAND and Borexino data indicate a preference for a greater contribution of $\Um$ geo-neutrinos and a reduced contribution of $\Th$ geo-neutrinos when incorporating the new fluxes.
For the KamLAND data, we obtain $N_{\mathrm{U}}=122\pm40$ and $N_{\mathrm{Th}}=48\pm28$, which represents an increase of approximately $9.9\%$ for the $\Um$ contribution and a decrease of $14.2\%$ for the $\Th$ contribution. This results in a lower Th/U mass ratio of $5.82\pm4.10$. 
For the Borexino data, the updated results are $N_{\mathrm{U}}=36\pm17$ and $N_{\mathrm{Th}}=16\pm11$, yielding a Th/U mass ratio of $6.57\pm7.71$. In this case, there is approximately a $20\%$ increase in the $\Um$ contribution and a $20\%$ decrease in the $\Th$ contribution.
In Figure~\ref{fig.analysis}, the red lines represent the $\Delta \chi^2$ profiles in the $N_{\rm{U}}$-$N_{\rm{Th}}$ plane for KamLAND and Borexino under the assumption of the new geo-neutrino fluxes.


\begin{table*}
    \centering
    \setlength{\tabcolsep}{2mm}{
    \begin{tabular}{cccccc}
    \hline
    \hline
    \multirow{2}{*}{Flux Model Inputs} & \multirow{2}{*}{Contribution} & \multicolumn{2}{c}{KamLAND} & \multicolumn{2}{c}{Borexino}\\
    & & Best-fit number & Flux & Best-fit number & Flux\\
         \hline
         \multirow{4}{*}{Enomoto flux}& Geo-neutrinos (Th/U fixed) & $176\pm29$& $33.5\pm5.5$\,[TNU] & $52\pm9$ & $46.5\pm8.1$\,[TNU]\\
         {}& Geo-neutrinos (Th/U free) & $167\pm31$& $31.9\pm5.9$\,[TNU] & $50\pm11$ & $44.7\pm10.1$\,[TNU]\\
         {}& $\Um$ & $111\pm40$ & $21.0\pm7.6$\,[TNU] & $30\pm16$ & $26.6\pm14.2$\,[TNU]\\
         {} & $\Th$ & $56\pm30$ &$10.9\pm5.8$\,[TNU] & $20\pm11$ & $18.1\pm10.0$\,[TNU]\\
         {}& Mantle neutrinos (Th/U fixed) & - & $5.3^{+6.2}_{-5.3}$\,[TNU] & - & $20.9^{+9.5}_{-8.7}$\,[TNU]\\
         {}& Mantle neutrinos (Th/U free) & - & $3.7^{+6.6}_{-3.7}$\,[TNU] & - & $19.1^{+11.3}_{-10.9}$\,[TNU]\\
         \hline
         \multirow{4}{*}{New flux model}& Geo-neutrinos (Th/U fixed) & $179\pm29$ & $34.0\pm5.5$\,[TNU]& $54\pm9$& $48.3\pm8.1$\,[TNU]\\
         {} & Geo-neutrinos (Th/U free) & $170\pm31$ & $32.4\pm5.9$\,[TNU]& $52\pm12$& $46.4\pm10.4$\,[TNU]\\
         {}& $\Um$ & $122\pm40$ & $23.1\pm7.6$\,[TNU]& $36\pm17$& $31.9\pm15.1$\,[TNU]\\
         {} & $\Th$ & $48\pm28$ & $9.3\pm5.5$\,[TNU] & $16\pm11$ &$14.5\pm10.0$\,[TNU]\\
         {}& Mantle neutrinos (Th/U fixed) & - & $7.2\pm6.2$\,[TNU] & - & $23.9^{+9.5}_{-8.7}$\,[TNU]\\
         {}& Mantle neutrinos (Th/U free) & - & $4.2^{+6.6}_{-4.2}$\,[TNU] & - & $20.8^{+11.3}_{-10.9}$\,[TNU]\\
    \hline
    \hline
    \end{tabular}
    \caption{The best-fit numbers of IBD events and the corresponding fluxes of geo-neutrinos from the total contributions of $\Um$ and $\Th$, as well as from their individual contributions, as observed in KamLAND and Borexino. Terrestrial Neutrino Units (TNU) are used to describe the fluxes converted from the best-fit IBD events. 
    The first section presents the results of KamLAND and Borexino using the Enomoto flux, fitted with configurations similar to those in Refs.~\cite{KamLAND:2022vbm} and \cite{Borexino:2019gps}, respectively. The second section provides results fitted based on the new geo-neutrino flux model.   
}
    \label{tab.fitted_results}}
\end{table*}

The geo-neutrino signals originating from the mantle can be derived by subtracting the expected signals from the crustal models, as detailed in Table~\ref{tab.crust}, from the total observed geo-neutrino signal. Firstly, we assume a fixed Th/U ratio~\cite{MCDONOUGH1995223}. When considering the Enomoto flux~\cite{Enomoto}, the mantle signal for KamLAND is estimated to be $5.3^{+6.2}_{-5.3}\,\mathrm{TNU}$, while for Borexino, it is $20.9^{+9.5}_{-8.7} \, \mathrm{TNU}$.
Incorporating the new flux model, the mantle signal for KamLAND increases to $7.2\pm6.2\,\mathrm{TNU}$, and for Borexino, it rises to $23.9^{+9.5}_{-8.7}\, \mathrm{TNU}$.
Furthermore, we estimate the mantle signals when allowing for a free Th/U ratio. Under these conditions, the mantle signal for KamLAND yields $3.7^{+6.6}_{-3.7}\,\mathrm{TNU}$, and for Borexino, it results in $19.1^{+11.3}_{-10.9} \, \mathrm{TNU}$ based on the Enomoto inputs~\cite{Enomoto}. With the new flux model, we obtain $4.2^{+6.6}_{-4.2}\,\mathrm{TNU}$ for KamLAND and $20.8^{+11.3}_{-10.9}\,\mathrm{TNU}$ for Borexino.
It is found that there exists a tension of approximately $1.5\sigma$ ($1.2\sigma$) in the estimation of mantle signals between KamLAND and Borexino, irrespective of whether the analysis is based on the Enomoto flux or the new flux model, and regardless of whether the Th/U ratio is treated as fixed or free. The introduction of the new flux model does not mitigate the discrepancies observed in the current results for mantle geo-neutrino signals, which suggests a need for further explanations and support from next-generation geo-neutrino experiments.


In conclusion, this work presents a novel calculation of geo-neutrino fluxes and energy spectra originating from the decay chains of $^{238}\mathrm{U}$ and $^{232}\mathrm{Th}$, which are the primary contributors to geo-neutrino signals. The new summation spectra are derived from the latest data available in the ENSDF nuclear database~\cite{ensdf_web}, which provides comprehensive documentation of the transition information for all beta decay branches included in our calculations. Moreover, we account for the analytical description of the single beta decay spectrum presented in Ref.~\cite{Li:2019quv}, which incorporates various shape factors for forbidden transitions as well as higher-order corrections.
Regarding the update of the nuclear database, the observed variations arise from adjustments in the $Q$-values of specific decay branches, particularly the transition from ${}^{234}\mathrm{Pa} \to {}^{234}\mathrm{U}$. 
Furthermore, in contrast to Ref.~\cite{Enomoto}, this study introduces a new calculation for the forbidden decays of $\mathrm{{}^{212, 214}{Bi}}$ and $\mathrm{{}^{210}{Tl}}$, which represents a significant improvement of the geo-neutrino energy spectra.
By utilizing the updated geo-neutrino flux inputs and the IBD cross section, we can evaluate the IBD yield, which can be regarded as the event rate recorded in geo-neutrino detection experiments. Our findings indicate that the IBD yields for $\U$ and $\Th$ geo-neutrinos have been underestimated using the previous geo-neutrino fluxes. 

We have also examined the impact of geo-neutrino fluxes on the analysis of geo-neutrino data. By utilizing the recently released datasets from the KamLAND~\cite{KamLAND:2022vbm} and Borexino~\cite{Borexino:2019gps} collaborations, we present our statistical analysis results using both the Enomoto fluxes and the newly calculated fluxes from this work. Our findings indicate that, in comparison to the results derived from the Enomoto fluxes, both KamLAND and Borexino data show a preference for a greater proportion of $\U$ geo-neutrinos and a reduced presence of $\Th$ geo-neutrinos when using the new geo-neutrino fluxes.
Regarding the mantle signal, the new flux model leads to an increase for both KamLAND and Borexino datasets.
Overall, this new flux model is anticipated to play a vital role in next-generation experiments, which are expected to yield more precise geo-neutrino measurements. Our study represents a significant advancement in geo-neutrino research, establishing a new benchmark for accuracy and reliability in the field.
\\
\\

\begin{acknowledgments}
The current work was supported by the National Natural Science Foundation of China under Grant Nos.~12075255 and 11835013. 
\end{acknowledgments}

\bibliographystyle{apsrev4-1}
\bibliography{ref}

\clearpage

\appendix
\section{The analytical method for beta transition}
\label{sec.beta_decay}

The geo-neutrino spectrum is calculated using a summation method, which includes the evaluation of the individual beta decay spectrum, and the application of the nuclear database.
In this section, we describe our approach for calculating the single beta decay branches, following the strategy outlined in Ref.~\cite{Li:2019quv}. In the subsequent section, we introduce the nuclear database employed in this work.

In evaluating single beta decay, we adopt a universal description as presented in Ref.~\cite{Fermi:1934hr}. The beta spectrum can be expressed as follows:
\begin{align}
    S_\beta (E_e)= & K p_e E_e (E_0 - E_e)^2 F(Z, E_e)  \nonumber\\
    &C(Z, E_e) \left[1+ \delta(Z,A,E_e) \right],
\end{align}
where $K$ is the normalization factor, $F(Z, E_e)$ is the Fermi function that describes the effect of the Coulomb field on the outgoing electron, and $C(Z, E_e)$ is the shape factor that accounts for the distortion in forbidden transitions compared to allowed transitions, which takes into consideration the energy and momentum dependence of the nuclear matrix elements. 
As for the additional higher-order corrections $\delta (Z,A,E_e)$ which can be decomposed into the radiative correction (RC) $\delta_{\rm{RC}}$~\cite{Sirlin:1967zza,Sirlin:2011wg}, finite size (FS) correction $\delta_{\rm{FS}}$~\cite{Hayes:2013wra, Wang:2016rqh} and weak magnetism (WM) correction $\delta_{\rm WM}$~\cite{Hayes:2013wra}.

Geo-neutrinos, which are produced from the beta decays of nuclear isotopes with long half-lives, typically originate from nuclei characterized by high atomic numbers of $Z$. This results in strong Coulomb potentials. Consequently, the Fermi function can be represented as follows:
\begin{align}
    F(Z, E_\nu) = 2(\gamma+1)(2E_\nu R)^{2(\gamma-1)}e^{\pi y}\left|\frac{\Gamma(\gamma+iy)}{\Gamma(2\gamma+1)} \right|^2 ,
    \label{eq.fermi_function}
\end{align}
where $\gamma=\sqrt{1-(\alpha Z)^2}$, and $y=\alpha Z E_e/p_e$ with $\alpha$ denoting the fine structure constant. $R$ is the nuclear radius which can be expressed as a function of the nucleon number $A$:
\begin{align}
    R = 1.121A^{1/3} + 2.426A^{-1/3} - 6.614A^{-1},
    \label{eq.nucleon_radius}
\end{align}
where the unit of $R$ is femtometers (fm).

The expressions for RC differ between the electron spectrum and the antineutrino spectrum, as outlined below:
\begin{align}
    \delta_{\mathrm{RC}}^e & = \frac{\alpha}{2\pi} g(E_e, E_0),\\ 
    \delta_{\mathrm{RC}}^{\bar{\nu}} & = \frac{\alpha}{2\pi} h(E_e, E_0),
    \label{eq.radiative_correction}
\end{align}
where $g(E_e, E_0)$ and $h(E_e, E_0)$ are further defined as follows~\cite{Sirlin:1967zza,Sirlin:2011wg}:
\begin{widetext}
\begin{align}
    g(E_e, E_0) = & 3 \ln\left(\frac{M_N}{m_e}\right)-\frac{3}{4}+4\left(\frac{\tanh^{-1}\beta}{\beta} -1\right)\left[\frac{E_0-E_e}{3E_e}-\frac{3}{2} + \ln{\frac{2\left(E_0-E_e\right)}{m_e}}\right]\nonumber\\
    &+\frac{\tanh^{-1}{\beta}}{\beta}\left[2(1+\beta^2)+\frac{(E_0-E_e)^2}{6E_e^2} -4\tanh^{-1}\beta\right] +\frac{4}{\beta}L\left(\frac{2\beta}{1+\beta}\right),\\
    h(\hat{E}, E_0) = & 3 \ln\left(\frac{M_N}{m_e}\right)+\frac{23}{4}+\frac{8}{\beta}L\left(\frac{2\beta}{1+\beta}\right) + 8\left(\frac{\tanh^{-1}\hat{\beta}}{\beta} -1\right) \ln\left({\frac{2\hat{E}\hat{\beta}}{m_e}}\right) \nonumber\\
    & + 4 \frac{\tanh^{-1}{\hat{\beta}}}{\hat{\beta}}\left(\frac{7+3\hat{\beta}^2}{8}-2\tanh^{-1}{\hat{\beta}} \right),
    \label{eq.radiative_correction2}
\end{align}
\end{widetext}
where $L(x)$ is denoted as $L(x)=\int_0^x {dt}/{t} \ln(1-t)$, $\beta=p_e/E_e$, $\hat{E}=E_0-E_{\bar{\nu}}$, $\hat{\beta}=\hat{p}/\hat{E}$ with
$\hat{p} =\sqrt{\hat{E}^{2}-m^{2}_e}$, $M_N$ and $m_e$ represent the masses of the nucleon and electron, respectively.

The FS correction to the Fermi function for the allowed Gamow-Teller (GT) beta transition, up to the order of $\alpha Z$, can be defined as follows~\cite{Hayes:2013wra,Wang:2016rqh}:
\begin{align}
    \delta_{\mathrm{FS}}=-\frac{3}{2}Z\alpha\left\langle r \right\rangle \left(E_e - \frac{E_\nu}{27}  + \frac{m_e^2}{3E_e})\right),
    \label{eq.FS_correction}
\end{align}
where $\left\langle r \right\rangle=(36/35)R$, assuming a uniform distribution of weak and charge densities across a radius $R$.
Considering that the FS correction is operator-dependent, it seems impossible to derive a general and accurate expression that is applicable to all transitions. Furthermore, a satisfactory FS correction for first-forbidden transitions has yet to be established. Consequently, the correction for the allowed GT beta decay, as outlined in Eq.~\eqref{eq.FS_correction}, will be applied to all transitions, including the forbidden ones.

\begin{table*}
    \centering
    \renewcommand{\arraystretch}{1.8}
    \begin{tabular}{cccc}
    \hline\hline
         \multirow{2}*{Classification} & \multirow{2}*{$\Delta J^{\pi}$} & {Shape factor $C(Z,E_e)$} & \multirow{2}*{WM corrections}\\
         ~ & ~ & Exact relativistic calculation & ~ \\
         \hline
         Allowed GT & $1^+$ & $1$ & $\frac{2}{3}\frac{\mu_\nu-1/2}{M_N g_A}\left(E_e\beta^2-E_\nu\right)$\\
         Nonunique $1^{\rm{st}}$ forbidden GT & $0^-$ & $E_\nu^2 + p_e^2\tilde{F}_{p_{1/2}}+2p_eE_\nu\tilde{F}_{sp_{1/2}}$ & $0$ \\
         Nonunique $1^{\rm{st}}$ forbidden GT & $1^-$ & $E_\nu^2 + \frac{2}{3}p_e^2\tilde{F}_{p_{1/2}} + \frac{1}{3}p_e^2\tilde{F}_{p_{3/2}}-\frac{4}{3}p_eE_\nu\tilde{F}_{sp_{1/2}}$ & $\frac{\mu_\nu-1/2}{M_N g_A}\frac{(E_e\beta^2-E_\nu)(p_e^2+E_\nu^2)+2\beta^2E_eE_\nu(E_\nu-E_e)/3}{p_e^2+E_\nu^2-4E_\nu E_e/3}$\\
         Unique $1^{\rm{st}}$ forbidden GT & $2^-$ & $E_\nu^2 + p_e^2\tilde{F}_{p_{3/2}}$ & $\frac{3}{5}\frac{\mu_\nu-1/2}{M_N g_A}\frac{(E_e\beta^2-E_\nu)(p_e^2+E_\nu^2)+2\beta^2E_eE_\nu(E_\nu-E_e)/3}{p_e^2+E_\nu^2}$ \\
    \hline\hline
    \end{tabular}
    \caption{The shape factors $C(Z, E_e)$ and WM corrections for allowed and first forbidden GT transitions.
    The shape factors are calculated using the exact relativistic calculation (ERC) of the Dirac wave function, as presented in the third column.}
    \label{tab.shape_factors}
\end{table*}
The WM correction arises from the interference between the magnetic moment distribution of the vector current and the spin distribution of the axial current~\cite{Hayes:2016qnu,Wang:2017htp}, indicating that different types of transitions exhibit distinct forms of WM corrections. We adopt the WM corrections from Ref.~\cite{Hayes:2013wra}, which are presented in the final column of Table~\ref{tab.shape_factors}, in which $\mu_\nu= 4.7$ represents the nucleon isovector magnetic moment, $M_N$ denotes the nucleon mass, and $g_A$ is the axial vector coupling constant.

Finally let us discuss the shape factor $C(Z, E_e)$ for forbidden transitions. In our analysis, we treat all forbidden transitions as first-forbidden GT transitions, focusing on three representative cases listed in the second, third, and fourth rows of Table~\ref{tab.shape_factors}: 
the nonunique first forbidden GT transition with $\Delta J^\pi=0^-$, the nonunique first forbidden GT transition with $\Delta J^\pi = 1^-$, and the unique first forbidden GT transition with $\Delta J^\pi = 2^-$, where $\Delta J$ and $\Delta \pi$ represent the changes in angular momentum and parity between the initial and final nuclei, respectively.
In Table~\ref{tab.shape_factors}, we present the results of the exact relativistic calculation (ERC) of the Dirac wave function~\cite{Stefanik:2017dbz},  in which several Fermi-like functions are introduced and can be expressed as follows~\cite{Stefanik:2017dbz}:
\begin{widetext}
\begin{align}
    F_0(E_e, R) & = 2/(1+\gamma)F(E_e, Z), \\
    \tilde{F}_{p3/2}(E_e, R) & \simeq F_1(E_e, Z)/F_0(E_e,Z), \\
    \tilde{F}_{p1/2}(E_e, R) & \simeq \left[\left(\frac{\alpha Z}{2} + \frac{E_eR}{3}\right)^2 + \left(\frac{m_eR}{3} \right)^2 -\frac{2m_e^2R}{3E_e}\left(\frac{\alpha Z}{2} + \frac{E_eR}{3} \right) \right]/j_1^2(p_eR), \\
    \tilde{F}_{sp1/2}(E_e, R) & \simeq \left[\left(\frac{\alpha Z}{2} + \frac{E_eR}{3}\right) + \frac{m_e^2 R}{3E_e} \right]/(j_0(p_eR) j_1(p_eR)), 
    \label{eq.fermi_functions}
\end{align}
\end{widetext}
where the spherical Bessel functions of $j_0(p_eR)$ and $j_1(p_eR)$ are utilized.

\section{The nuclear database}
\label{sec.database}

In this section, we provide details about the Evaluated Nuclear Structure Data Files (ENSDF) database~\cite{ensdf_web} concerning the decay chains of $\Um$ and $\Th$. Considering the detection capabilities of liquid scintillator detectors, only geo-neutrinos from the decay chains of $\Um$ and $\Th$ can be observed. For the decay chain of $\Um$ $\to$ ${}^{206}\mathrm{Pb}$, there are eight alpha decays and six beta decays; the beta decay information for this chain is listed in Table~\ref{tab.beta_decay_in_238U}. In contrast, the decay chain from $\Th$ to ${}^{208}\mathrm{Pb}$ includes six alpha decays and four beta decays, with the relevant beta decay information presented in Table~\ref{tab.beta_decay_in_232Th}.
In both tables, the first column presents the parent and daughter nuclei for each beta transition. The second column records $R_{ij}$, which represents the weight of the production ratio in the decay from isotope $i$ to isotope $j$. The corresponding $Q$-value for each transition is listed in the third column. Additionally, for each transition, we count the total number of transitions and the effective number of transitions that can be detected by the IBD reaction in the fourth column.


\begin{table*}
    \centering
    \setlength{\tabcolsep}{12mm}{
    \begin{tabular}{cccc}
    \hline\hline
         $i$ $\to$ $j$ & $R_{ij}$ & $Q$-value [keV] & transition number\\
         \hline
         ${}^{234}\rm{Th} \to {}^{234}\rm{Pa}$ &  $1.0000$ & $199.5$ & 5 (0) \\
         ${}^{234}\rm{Pa}^m \to {}^{234}\rm{U}$ & {$0.9984$} & $2290.0$ & 24 (1) \\
         ${}^{214}\rm{Pb} \to {}^{214}\rm{Bi}$ &  $0.9998$ & $1018.0$ & 7 (0) \\
         ${}^{214}\rm{Bi} \to {}^{214}\rm{Po}$ &  $0.9998$ & $3269.0$ & 70 (6) \\
         ${}^{210}\rm{Pb} \to {}^{210}\rm{Bi}$ &  $1.0000$ & $63.5$ & 2 (0) \\
         ${}^{210}\rm{Bi} \to {}^{210}\rm{Po}$ &  {$0.9999$} & $1161.5$ & 1 (0) \\
         ${}^{234}\rm{Pa} \to {}^{234}\rm{U}$  &  {$0.0016$} & $1247$ & 39 (0) \\
         ${}^{218}\rm{Po} \to {}^{218}\rm{At}$ &  $0.0002$ & $264.0$ & 1 (0) \\
         ${}^{206}\rm{Tl} \to {}^{206}\rm{Pb}$ &  {$0.0001$} & $1532.3$ & 3 (0) \\
         ${}^{210}\rm{Tl} \to {}^{210}\rm{Pb}$ &  $0.0002$ & $4386.0$ & 7 (5) \\
    \hline\hline
    \end{tabular}
    \caption{
    Beta decay transitions in the $\Um$ chain. For each transition, the weight of the production ratio $R_{ij}$, the $Q$-value, and the number of decay branches are provided. The last column lists the total number of decay branches as well as the effective decay branches that can be detected by the IBD reaction.
    }
    \label{tab.beta_decay_in_238U}}
\end{table*}

\begin{table*}
    \centering
    \setlength{\tabcolsep}{12mm}{
    \begin{tabular}{cccc}
    \hline\hline
         $i$ $\to$ $j$ & $R_{ij}$ & $Q$-value [keV] & transition number\\
         \hline
         ${}^{228}\rm{Ra} \to {}^{228}\rm{Ac}$ &  $1.0000$ & $39.5$ & 4 (0) \\
         ${}^{228}\rm{Ac} \to {}^{228}\rm{Th}$ & $1.0000$ & $2076.0$ & 55 (2) \\
         ${}^{212}\rm{Pb} \to {}^{212}\rm{Bi}$ &  $1.0000$ & $569.1$ & 3 (0) \\
         ${}^{212}\rm{Bi} \to {}^{212}\rm{Po}$ &  $0.6406$ & $2251.5$ & 7 (1) \\
         ${}^{208}\rm{Tl} \to {}^{208}\rm{Pb}$ &  $0.3594$ & $1801.3$ & 15 (0) \\
    \hline\hline
    \end{tabular}
    \caption{
    Beta decay transitions in the $\Th$ chain. For each transition, the weight of the production ratio $R_{ij}$, the $Q$-value, and the number of decay branches are provided. The last column lists the total number of decay branches as well as the effective decay branches that can be detected by the IBD reaction. 
    }
    \label{tab.beta_decay_in_232Th}}
\end{table*}
\begin{table*}
    \centering
    \setlength{\tabcolsep}{4mm}{
    \begin{tabular}{c|cccccc}
    \hline\hline
         decay chains & $i \to j$ & $R_{i,j}$ & $Q$-value [keV] & $I_{ij,k}$ & $\delta I_{ij,k}$ & Transition type  \\
         \hline
         \multirow{5}*{$\Um$} & ${}^{234}\rm{Pa}^{\rm{m}}\to{}^{234}\rm{U}$ & $1.0000$ & $2290.0$ & $0.9757$ & $0.0004$ & $1^{\rm st} \, \text{forbidden} \, (0^- \to 0^+)$ \\
         \cline{2-7}
         ~ & \multirow{6}*{${}^{214}\rm{Bi}\to{}^{214}\rm{Po}$} & \multirow{6}*{$0.9998$} & $3269.0$ & $0.192$ & $0.004$ & $1^{\rm st}\text{forbidden} \, (1^- \to 0^+)$ \\
         ~ & ~ & ~ & $2660.0$ & $0.0055$ & $0.0008$ & $1^{\rm st} \, \text{forbidden} \, (1^- \to 2^+)$ \\
         ~ & ~ & ~ & $2254.0$ & $0.00079$ & $0.00013$ & $3^{\rm rd} \, \text{forbidden} \, (1^- \to 4^+)$ \\
         ~ & ~ & ~ & $1994.0$ & $0.0006$ & $0.0004$ & $2^{\rm nd} \, \text{forbidden} \, (1^- \to 3^-)$ \\
         ~ & ~ & ~ & $1891.0$ & $0.0722$ & $0.0008$ & $1^{\rm st} \, \text{forbidden} \, (1^- \to 2^+)$ \\
         ~ & ~ & ~ & $1854.0$ & $0.009$ & $0.0005$ & $1^{\rm st} \, \text{forbidden} \, (1^- \to 0^+)$ \\
         \cline{2-7}
         ~ & \multirow{5}*{${}^{210}\rm{Tl}\to{}^{210}\rm{Pb}$} & \multirow{5}*{$0.0002$} & $4386.0$ & $0.20$ & $\text{Unknown}$ & $\text{Allowed} \, (5^+ \to 4^+)$ \\
         ~ & ~ & ~ & $4210.0$ & $0.30$ & $0.06$ & $2^{\rm rd} \, \text{forbidden} \, (5^+ \to 8^+)$ \\
         ~ & ~ & ~ & $2413.0$ & $0.10$ & $0.03$ & $2^{\rm rd} \, \text{forbidden} \, (5^+ \to 2^+)$ \\
         ~ & ~ & ~ & $2020.0$ & $0.10$ & $0.03$ & $\text{Allowed}\, (5^+ \to 4^+)$ \\
         ~ & ~ & ~ & $1860.0$ & $0.24$ & $0.05$ & $\text{Unknown}$ \\
         \hline
         \multirow{3}*{$\Th$} & ${}^{212}\rm{Bi}\to{}^{212}\rm{Po}$ & $0.6406$ & $2251.5$ & $0.8643$ & $0.0012$ & $1^{\rm st} \, \text{forbidden} \, (1^- \to 0^+)$ \\
         \cline{2-7}
         ~ & \multirow{2}*{${}^{228}\rm{Ac}\to{}^{228}\rm{Th}$} & \multirow{2}*{$1.0000$} & $2076.0$ & $0.07$ & $0.05$ & $\text{Allowed} \, (3^+ \to 2^+)$ \\
         ~ & ~ & ~ & $1947.0$ & $0.006$ & $0.005$ & $\text{Allowed} \, (3^+ \to 4^+)$ \\
    \hline\hline
    \end{tabular}}
    \caption{Effective transitions above the IBD threshold in the decay chains of $\Um$ and $\Th$. In addition to the $R_{ij}$ and $Q$-values provided in Table~\ref{tab.beta_decay_in_238U} and~\ref{tab.beta_decay_in_232Th}, the intensity $I_{ij,k}$, its uncertainty $\delta I_{ij,k}$, and the transition type are also listed. These values are obtained from the latest ENSDF nuclear database~\cite{ensdf_web}.}
    \label{tab.forbidden_decay}
\end{table*}

Considering the threshold energy of the IBD reaction with $E^{\mathrm{thr}}_\nu=1.806\,\mathrm{MeV}$, only three nuclides in the decay chain of $\Um$ (i.e., ${}^{234}\mathrm{Pa}^{\mathrm{m}}$, ${}^{214}\mathrm{Bi}$, ${}^{210}\mathrm{Tl}$) 
generate detectable antineutrinos. Among these, only 12 decay branches contribute to the observed geo-neutrino signal, as listed in Table~\ref{tab.forbidden_decay}.
Notably, there is one branch for which the associated decay transition cannot be clearly identified. Since its contribution is negligible, we categorize this branch as an allowed transition for simplicity. Consequently, the total viable branches comprises nine forbidden branches and three allowed branches.
For the first forbidden transitions and the allowed transitions, the calculations are detailed in Appendix~\ref{sec.beta_decay}.
Regarding the second and third forbidden transitions, in the absence of exact relativistic calculations, we first apply shape factors derived from previous studies~\cite{Mueller:2011nm}. Additionally, we utilize an approximation that treats the second and third forbidden transitions as allowed transitions. The spectra obtained from these two methods have a negligible impact on our results. In this work, Figure~\ref{fig.spectra} illustrates the calculations for the second and third forbidden transitions, treated in the same manner as the allowed transitions.

In the decay chain of $\Th$, the contributions of observable geo-neutrinos come solely from two nuclides (i.e., ${}^{212}{\mathrm{Bi}}$ and ${}^{228}{\mathrm{Th}}$), corresponding to three beta decay transitions. Among these three transitions, there is only one non-unique first forbidden transition, while the remaining two transitions are classified as allowed ones. The decay information for all transitions in the decay chain of $\Th$ is summarized in Table~\ref{tab.beta_decay_in_232Th}, which includes the three effective transitions that exceed the IBD threshold, as listed in Table~\ref{tab.forbidden_decay}.

To summarize, based on the descriptions in Appendix~\ref{sec.beta_decay} and~\ref{sec.database}, we can obtain the total geo-neutrino spectrum by utilizing the latest ENSDF nuclear database and the energy spectrum calculation for individual decay branches:
\begin{align}
    S_{X} = \sum_{ij} R_{ij} \sum_k I_{ij,k} S_\nu^{ij,k},
\end{align}
where $R_{ij}$ and $I_{ij,k}$ are taken directly from Tables~\ref{tab.beta_decay_in_238U} through~\ref{tab.forbidden_decay}, while $S_\nu^{ij,k}$ is the single beta decay spectrum described in Appendix A.

\section{The geo-neutrino analysis}
\label{sec.geo_analysis}

In this section, we describe our replication of the geo-neutrino analysis based on data released by KamLAND~\cite{KamLAND:2022vbm} and Borexino~\cite{Borexino:2019gps} collaborations. 
The least squares method is employed, using a binned $\chi^2$ that includes both the statistical part and additional pull terms that account for the systematic uncertainties. Earlier reproductions of the geo-neutrino analyses from KamLAND and Borexino data can be found in Refs.~\cite{Fogli:2010vx,Fiorentini:2012yk}.

\subsection{Geo-neutrino analysis of KamLAND data}

\begin{figure*}
\centering
\makebox[\textwidth][c]{\includegraphics[width=1.2\textwidth]{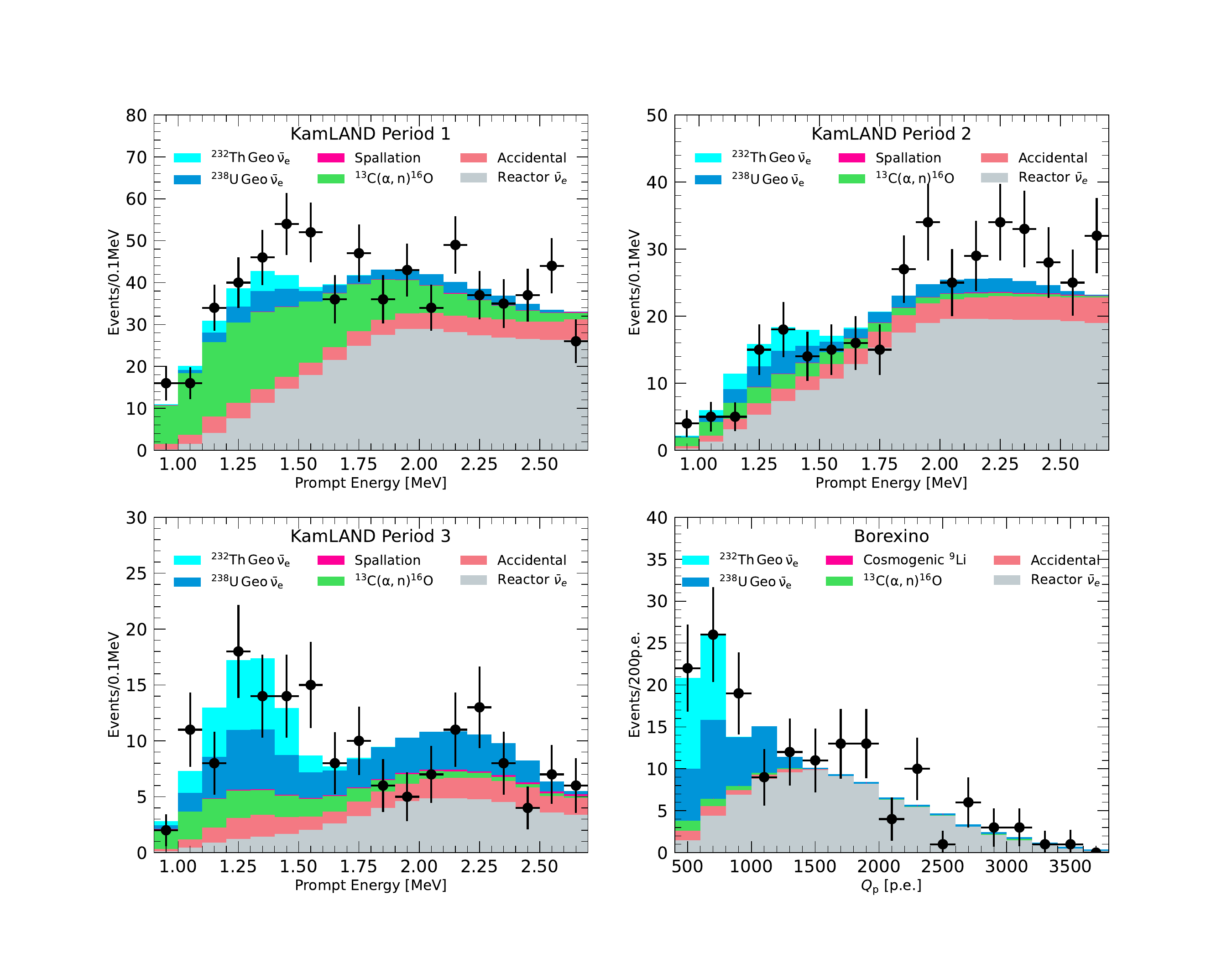}}
\caption{The observed event spectra (data points) and the best fits (histograms) for different components of the geo-neutrino signal and backgrounds are presented for KamLAND (across three periods) and Borexino, based on Eq.~\eqref{eq.chi2_kamland1} and Eq.~\eqref{eq.chi2_borexino}, respectively. Note that the contributions from U and Th are treated as free parameters.}
\label{fig.best_fit}
\end{figure*}

In our analysis of the KamLAND data, we referred to the observations presented in Ref.~\cite{KamLAND:2022vbm}, which are divided into three distinct periods. Given that the detector responses vary across these periods, we structure our total $\chi^2$ as the sum of three separate $\chi^2_j$ functions ($j = \mathrm{P_{1}}, \mathrm{P_{2}}, \mathrm{and}\; \mathrm{P_{3}}$), corresponding to Period 1, Period 2, and Period 3, respectively. Additionally, we include extra pull terms to account for systematic uncertainties:
\begin{widetext}
\begin{align}
    & \chi^{2} =  \chi^2_{\mathrm{P_{1}}} + \chi^2_{\mathrm{P_{2}}} + \chi^2_{\mathrm{P_{3}}} + \mathrm{pull \, terms} , \label{eq.chi2_kamland1} \\
    & \chi^{2}_j = \sum_i^{\mathrm{Nbins}} \frac{\left(M_{ij} - \left(\left(1+\alpha_j^{\mathrm{Rea}}\right)T_{ij}^{\mathrm{Rea}} + T_{ij}^{\mathrm{Geo}} + \sum_{\mathrm{B}}^{\mathrm{Bkg}} (1+\alpha_j^{\mathrm{B}})T_{ij}^{\mathrm{B}}\right)\right)^2}{(\delta M_{ij})^2} ,  \label{eq.chi2_kamland2}\\
    & \mathrm{pull \, terms} = \sum_j^3 \left(\frac{\alpha_j^{\mathrm{Rea}}}{\sigma_j^{\mathrm{Rea}}} \right)^2 + \sum_j^3\sum_{\mathrm{B}}^{\mathrm{Bkg}}\left(\frac{\alpha_j^{\mathrm{B}}}{\sigma_j^{\mathrm{B}}} \right)^2\;,
\label{eq.chi2_kamland3}
\end{align}
\end{widetext}
where the spectral data of the observed events $M_{ij}$ for these three periods are obtained from Ref.~\cite{KamLAND:2013rgu}, and are presented as data points in Figure~\ref{fig.best_fit}.
The associated uncertainty $\delta M_{ij}$ encompasses both statistical uncertainties and uncorrelated bin-to-bin shape uncertainties, which are also illustrated in Figure~\ref{fig.best_fit}.
The overall uncertainties from reactors are specified as $\sigma_j^{\mathrm{Rea}} = 3.5\%$ for Period 1 and $\sigma_j^{\mathrm{Rea}}=4.0\%$ for Periods 2 and 3. 
The background uncertainties, denoted as $\sigma_{\mathrm{B}}$', are derived from Table 1 of Ref.~\cite{KamLAND:2013rgu}, and are also reproduced in Table~\ref{tab.bkg_in_KamLAND} of this section.

\begin{center}
\begin{table*}[t]
\centering
\begin{tabular}{@{}*{22}{llr@{}l@{}c@{}r@{}lr@{}l@{}c@{}r@{}lr@{}l@{}c@{}r@{}lr@{}l@{}c@{}r@{}l}}
\hline\hline
\multicolumn{2}{l}{Background}  & \multicolumn{5}{c}{~~Period 1~~~~~~~~~~~} & \multicolumn{5}{c}{Period 2~~~~~~~~~~~} & \multicolumn{5}{c}{Period 3~~~~~~~~~~~} & \multicolumn{5}{c}{~~All Periods~~~~~} \\
 \multicolumn{2}{l}{} & \multicolumn{5}{c}{~~(1486 days)~~~~~~~~~~~} & \multicolumn{5}{c}{(1154 days)~~~~~~~~~~~} & \multicolumn{5}{c}{(351 days)~~~~~~~~~~~} & \multicolumn{5}{c}{~~(2991 days)~~~~~} \\
\hline
1 & Accidental & 76.&1 & ~$\pm$~ & 0.&1 & 44.&7 & ~$\pm$~ & 0.&1 & 4.&7 & ~$\pm$~ & 0.&1 & 125.&5 & ~$\pm$~ & 0.&1\\
2 & $^{9}$Li/$^{8}$He & 17.&9 & $\pm$ & 1.&4 & 11.&2 & $\pm$ & 1.&1 & 2.&5 & $\pm$ & 0.&5 & 31.&6 & $\pm$ & 1.&9\\
\multirow{2}{*}{3 $\bigg{\lbrace}$}  
& $^{13}{\rm C}(\alpha,{\it n})^{16}{\rm O}_{\rm g.s.}$, elastic scattering & 160.&4 & ~$\pm$~ & 16.&4 & 16.&5 & ~$\pm$~ & 3.&8 & 2.&3 & ~$\pm$~ & 1.&0 & 179.&0 & ~$\pm$~ & 21.&1\\
& $^{13}{\rm C}(\alpha,{\it n})^{16}{\rm O}_{\rm g.s.}$, $^{12}{\rm C}({\it n},{\it n'})^{12}{\rm C}^{*}$ (4.4 MeV $\gamma$) & 6.&9 & ~$\pm$~ & 0.&7 & 0.&7 & ~$\pm$~ & 0.&2 & 0.&10 & ~$\pm$~ & 0.&04 & 7.&7 & ~$\pm$~ & 0.&9\\
\multirow{2}{*}{4 $\bigg{\lbrace}$} 
& $^{13}{\rm C}(\alpha,{\it n})^{16}{\rm O^{*}}$, 1st e.s. (6.05 MeV $e^{+}e^{-}$) & 14.&6 & ~$\pm$~ & 2.&9 & 1.&7 & ~$\pm$~ & 0.&5 & 0.&21 & ~$\pm$~ & 0.&09 & 16.&5 & ~$\pm$~ & 3.&5\\
& $^{13}{\rm C}(\alpha,{\it n})^{16}{\rm O^{*}}$, 2nd e.s. (6.13 MeV $\gamma$) & 3.&4 & ~$\pm$~ & 0.&7 & 0.&4 & ~$\pm$~ & 0.&1 & 0.&05 & ~$\pm$~ & 0.&02 & 3.&9 & ~$\pm$~ & 0.&8\\
5 & Fast neutron and atmospheric neutrinos~~~~~ & \multicolumn{5}{c}{$<$ 7.7}~~~~~~~ & \multicolumn{5}{c}{$<$ 5.9}~~~~~~~ & \multicolumn{5}{c}{$<$ 1.7}~~~~~~~~~ & \multicolumn{5}{c}{$<$ 15.3}~~\\
\hline
Total  \hspace{-0.3cm} &  & 279.&2 & $\pm$ & 22.&1 & 75.&2 & $\pm$ & 7.&6 & 9.&9 & $\pm$ & 2.&1 & 364.&1 & $\pm$ & 30.&5\\
\hline\hline
\end{tabular}
\caption{\label{tab.bkg_in_KamLAND}Estimated backgrounds for $\overline{\nu}_{e}$ in the energy range between $0.9\,{\rm MeV}$ and $8.5\,{\rm MeV}$ after event selection cuts for KamLAND, which is presented in Ref.~\cite{KamLAND:2022vbm}.
}
\end{table*}
\end{center}

The predicted reactor neutrino signals for the $i$-th energy bin and $j$-th period $T_{ij}^{\mathrm{Rea}}$ and the corresponding backgrounds $T_{ij}^{\mathrm{B}}$ are taken from Ref.~\cite{KamLAND:2022vbm}, and reproduced in Table~\ref{tab.bkg_in_KamLAND} and Figure~\ref{fig.best_fit}.
Note that the backgrounds listed in Table~\ref{tab.bkg_in_KamLAND} are slightly different from the background components released in Ref.~\cite{KamLAND:2022vbm}, where we have ignored the backgrounds of fast neutrons and atmospheric neutrinos. 
Finally, the predictions for geo-neutrinos are derived as follows:
\begin{align}
    T_{ij}^{\mathrm{Geo}} = N_j^{\mathrm{U}}\cdot \epsilon_{ij} \cdot D_i^{\mathrm{U}}(E_{\mathrm{prompt}}) + N_j^{\mathrm{Th}} \cdot\epsilon_{ij}\cdot D_i^{\mathrm{Th}}(E_{\mathrm{prompt}})
\end{align}
where $N_j^{\mathrm{U}}$ and $N_j^{\mathrm{Th}}$ are the numbers of geo-neutrino signals, which are free parameters to be fitted in our $\chi^2$ function mentioned in Eqs.~\eqref{eq.chi2_kamland1}-\eqref{eq.chi2_kamland3}. $\epsilon_{ij}$ is the energy-dependent detection efficiency of the $j$-th period.
The predictions of the normalized energy spectra of geo-neutrino signals
$D_i^{\mathrm{U}}(E_{\mathrm{prompt}})$ and $D_i^{\mathrm{Th}}(E_{\mathrm{prompt}})$ are constructed from a shift from the neutrino energy to the deposited energy and then convoluted with the detector energy resolution.  
Furthermore, the observable neutrino spectra $D_i^{\mathrm{U}}(E_\nu)$ and $D_i^{\mathrm{Th}}(E_\nu)$ are calculated from the geo-neutrino energy spectra $S_i(E_\nu)$ and IBD cross section $\sigma_{\mathrm{IBD}}(E_\nu)$:
\begin{align}
    D^{\mathrm{U}}(E_\nu) & = S^{\mathrm{U}}(E_\nu) \cdot \sigma_{\mathrm{IBD}}(E_\nu)\;,\\
    D^{\mathrm{Th}}(E_\nu) & = S^{\mathrm{Th}}(E_\nu) \cdot \sigma_{\mathrm{IBD}}(E_\nu)\;.
\end{align}

\subsection{The geo-neutrino analysis for Borexino data}
\label{subsec.geo_borexino}

\begin{table*}
	\centering
        \setlength{\tabcolsep}{12mm}{
	\begin{tabular}{cc}
	\hline\hline
		Background Type & Events   \\
		\hline
	    $^9$Li background& 3.6 $\pm$ 1.0  \\
		Untagged muons & 0.023 $\pm$ 0.007 \\
		Fast n's ($\mu$ in WT) & $<$0.013 \\
		Fast n's ($\mu$ in rock) & $<$1.43  \\
		Accidental coincidences & 3.846 $\pm$ 0.017 \\
		($\alpha$, n) in scintillator & 0.81 $\pm$ 0.13 \\ 
		($\alpha$, n) in buffer & $<$2.6  \\
		($\gamma$, n) & $<$0.34 \\
		Fission in PMTs & $<$0.057   \\
		$^{214}$Bi-$^{214}$Po & 0.003 $\pm$ 0.001 \\
		\hline 
		Total & 8.28 $\pm$ 1.01  \\			
	\hline\hline
	\end{tabular}}	
        \caption{Summary of the expected number of events from non-antineutrino backgrounds in the antineutrino candidate sample in Borexino, which are taken from Ref.~\cite{Borexino:2019gps}. } 
        \label{tab.bkg_in_borexino}
\end{table*}
For the Borexino data, we utilized the dataset released in Ref.~\cite{Borexino:2019gps}, which corresponds to an exposure of 3262.74 days. The $\chi^{2}$ function is constructed in a manner similar to that used for the KamLAND data, with the only difference being the absence of the $j$ index:
\begin{widetext}
\begin{align}
    & \chi^{2} = \sum_i^{\mathrm{Nbins}} \frac{\left(M_{i} - \left(\left(1+\alpha^{\mathrm{Rea}}\right)T_{i}^{\mathrm{Rea}} + T_{i}^{\mathrm{Geo}} + \sum_{\mathrm{B}}^{\mathrm{Bkg}} (1+\alpha^{\mathrm{B}})T_{i}^{\mathrm{B}}\right)\right)^2}{\Delta M_i^2} + \left(\frac{\alpha_j^{\mathrm{Rea}}}{\sigma_j^{\mathrm{Rea}}} \right)^2 + \sum_{\mathrm{B}}^{\mathrm{Bkg}}\left(\frac{\alpha_j^{\mathrm{B}}}{\sigma^{\mathrm{B}}} \right)^2 , 
    \label{eq.chi2_borexino}
\end{align}
\end{widetext}
where $T_i^{\mathrm{Rea}}$ represents the prediction of the observed reactor neutrino spectrum, which is obtained from Figure 32(c) of Ref.~\cite{Borexino:2019gps}. The background contributions are referenced from Table 15 of the same publication and are summarized in Table~\ref{tab.bkg_in_borexino}.

\subsection{The mantle study}

The mantle geo-neutrino signal is derived by subtracting the expected signals from the crustal model from the total signal. The reference crustal contributions for Kamioka and Gran Sasso, used for KamLAND and Borexino, respectively, are detailed in Table~\ref{tab.crust}.
Regarding the uncertainties of the crustal model, we adopt the estimates provided in Ref.~\cite{Borexino:2019gps} for Borexino, while for KamLAND, we assume a 10\% uncertainty based on Ref.~\cite{Takeuchi:2019fft}.
It is important to note that the reference crustal model is calculated based on the Enomoto flux. Therefore, if we incorporate the new flux model as the spectral inputs, we must take into account the relative differences for $\Um$ and $\Th$ as listed in Table~\ref{tab.ibd_yield}.

\begin{table*}
    \centering
    \setlength{\tabcolsep}{4mm}{
    \begin{tabular}{ccc}
    \hline
    \hline
    {Flux Model Inputs} &  {KamLAND~\cite{KamLAND:2022vbm}} &{Borexino~\cite{Borexino:2019gps}}\\
         \hline
         {Enomoto flux}&  $28.2\pm2.9$ [TNU] & $25.6^{+5.0}_{-4.0}$ [TNU]\\
         \hline
         {The new flux}&  $26.8\pm2.7$ [TNU]& $24.3^{+4.8}_{-3.8}$ [TNU]\\
    \hline
    \hline
    \end{tabular}
    \caption{\label{tab.crust}Reference crustal contributions at Kamioka and Gran Sasso for KamLAND and Borexino, respectively. As for the uncertainties of the crustal model, we take the estimation in Ref.~\cite{Borexino:2019gps} for Borexino, while we assume $10\%$ for KamLAND~\cite{Takeuchi:2019fft}.  }}
\end{table*}

\end{document}